\documentclass[12pt,thmsa]{article}
\usepackage{amssymb}

\usepackage{sw20lart}
\usepackage{graphics}
\usepackage{epsfig}
\usepackage{sw20lart}

\input tcilatex

\begin{document}

\author{John D. Barrow\thanks{%
e-mail: \texttt{J.D.Barrow@damtp.cam.ac.uk}} \ and \ David F. Mota\thanks{%
e-mail: \texttt{D.F.Mota@damtp.cam.ac.uk}} \\
{\normalsize \textsl{Department of Applied Mathematics and Theoretical
Physics}}\\
{\normalsize \textsl{Centre for Mathematical Sciences, University of
Cambridge}}\\
{\normalsize \textsl{Wilberforce Road, Cambridge CB3 0WA, UK}}}
\title{Qualitative Analysis of Universes with Varying Alpha}
\date{}
\maketitle

\begin{abstract}
Assuming a Friedmann universe which evolves with a power-law scale
factor, $a=t^{n}$, we analyse the phase space of the system of
equations that 
describes a time-varying fine structure 'constant', $\alpha$, in the
Bekenstein-Sandvik-Barrow-Magueijo generalisation of general relativity. We
have classified all the possible behaviours of $\alpha (t)$ in
ever-expanding universes with different $n$ and find new exact solutions for 
$\alpha (t)$. We find the attractors points in the phase space for all $n$.
In general, $\alpha $ will be a non-decreasing function of time that
increases logarithmically in time during a period when the expansion is dust
dominated ($n=2/3$), but becomes constant when $n>2/3$. This includes the
case of negative-curvature domination ($n=1$). $\alpha $ also tends rapidly
to a constant when the expansion scale factor increases exponentially. A
general set of conditions is established for $\alpha $ to become
asymptotically constant at late times in an expanding universe.
\end{abstract}

\section{Introduction}

Stimulated by the observations for small variations in atomic structure
controlled by the fine structure constant in quasar absorption lines at
redshifts $z=1-3$, \cite{murphy,webb2,webb}, there has been much recent
interest in the theoretical predictions of gravity theories which extend
general relativity to incorporate space-time variations of the fine
structure 'constant'. These have been primary formulated as Lagrangian
theories with explicit variation of the velocity of light, $c$, \cite
{moffatal,am,ba}, or of the charge on the electron, $e,$ \cite
{bsbm,bsm1,bsm2,bsm3}. Theories of the latter sort offer the possibility of
matching the magnitude and trend of the quasar observations and have been
studied numerically and by means of matched asymptotic approximations. They
are also of particular interest because they predict that violations of the
weak equivalence principle should be observed at a level that is within
about an order of magnitude of existing experimental bounds \cite
{bsm4,zal,mof}. They are consistent with all other astrophysical and
experimental limits of time variation of the fine structure constant and
predict effects of the microwave background radiation, primordial
nucleosynthesis, and the Oklo natural reactor that are too small to conflict
with current observational bounds. A range of variant theories have been
investigated with attention to the possible particle physics motivations and
consequences for systems of grand and partial unification in references \cite
{banks,guts,olive}. In this paper we will give a full qualitative analysis
of the properties of Friedmann cosmological models in a sub-class of these
theories developed initially by Bekenstein \cite{bek2} to generalise
Maxwell's equations to include varying $e$ and then generalised by Sandvik,
Barrow and Magueijo \cite{bsbm} to include gravitation. We refer to these as
BSBM theories. We provide a phase-space analysis of the non-linear
propagation equation for the scalar field which carries the variations of
the fine structure constant. Some new exact solutions are also given and all
the asymptotic behaviours classified.

\section{The BSBM Theory}

\subsection{The Cosmological Evolution Equations}

We will assume that the total action of the Universe is given by BSBM:

\begin{equation}
S=\int d^4x\sqrt{-g}\left( \mathcal{L}_{grav}+\mathcal{L}_{matter}+\mathcal{L%
}_\psi +\mathcal{L}_{em}e^{-2\psi }\right) ,  \label{S}
\end{equation}
In the BSBM varying $\alpha $ theory, the quantities $c$ and $\hbar $ are
taken to be constant, while $e$ varies as a function of a real scalar field $%
\psi ,$ with

\begin{equation}
e=e_0e^\psi  \label{e}
\end{equation}
where $\mathcal{L}_\psi ={\frac \omega 2}\partial _\mu \psi \partial ^\mu
\psi $, $\omega $ is a coupling constant, and $\mathcal{L}_{em}=-\frac
14f_{\mu \nu }f^{\mu \nu }$. The gravitational Lagrangian is the usual $%
\mathcal{L}_g=-\frac 1{16\pi G}R$, with $R$ the curvature scalar and we have
defined an auxiliary gauge potential $a_\mu =\epsilon A_\mu ,$ and field
tensor $f_{\mu \nu }=\epsilon F_{\mu \nu }=\partial _\mu a_\nu -\partial
_\nu a_\mu $, so the covariant derivative takes the usual form, $D_\mu
=\partial _\mu +ie_0a_\mu $. The dependence on $\epsilon $ in the Lagrangian
then occurs only in the kinetic term for $\epsilon $ and in the $%
F^2=f^2/\epsilon ^2$ term.

The universe will be described by a homogeneous and isotropic Friedmann
metric with expansion scale factor $a(t)$ and curvature parameter $k.$
Varying the total Lagrangian we obtain the Friedmann equation ($G=c\equiv 1$%
) for a universe containing pressure-free matter and radiation: 
\begin{equation}
\left( \frac{\dot{a}}a\right) ^2=\frac{8\pi }3\left( \rho _m\left( 1+\left|
\zeta \right| e^{-2\psi }\right) +\rho _re^{-2\psi }+\rho _\psi +\rho
_\Lambda \right) -\frac k{a^2}  \label{fried}
\end{equation}
where the cosmological vacuum energy $\rho _\Lambda $ is a constant given by 
$\Lambda /(8\pi )$, and $\rho _\psi =\frac \omega 2\dot{\psi}^2$ and $\zeta =%
\mathcal{L}_{em}/\rho _m$ is the fraction of the matter which carries
electric or magnetic charge.

For the scalar field we obtain the evolution equation 
\begin{equation}
\ddot{\psi}+3H\dot{\psi}=-\frac 2\omega \exp [-2\psi ]\zeta \rho _m
\label{psidot}
\end{equation}
where $H\equiv \dot{a}/a$ is the Hubble parameter$.$ The conservation
equations for the non-interacting radiation and matter densities, $\rho _r$
and $\rho _m$ respectively, are: 
\begin{eqnarray}
\dot{\rho _m}+3H\rho _m &=&0  \label{mat} \\
\dot{\rho}_r+4H\rho _r &=&2\dot{\psi}\rho _r.  \label{dotrho}
\end{eqnarray}
so $\rho _m\propto a^{-3}$. However, the last relation can be written as 
$\dot{\tilde{\rho}_r}+4H\tilde{\rho}_r=0,$ 
with $\tilde{\rho}_r\equiv \rho _re^{-2\psi }\propto a^{-4}$. 

The Friedmann models with varying $\alpha $ have been shown \cite{bsm1} to
have the property that when $\zeta <0$ the homogeneous motion of the $\psi $
does not create significant metric perturbations at late times. This means
that far from the singularity we can safely assume that the expansion scale
factor follows the form for the Friedmann universe containing the same fluid
when $\alpha $ does not vary $(\zeta =0=\psi )$. The behaviour of $\psi $
then follows from a solution of equation (\ref{psidot}) in which $a(t)$ has
the form for a Friedmann universe for matter with the same equation of state
in general relativity when $\zeta =0=\psi .$ This behaviour is natural. We
would not expect that very small variations in the coupling to
electromagnetically interacting matter would have large gravitational
effects upon the expansion of the universe. Thus, in this paper we will
provide a complete analysis of the behaviour of the solutions of the
non-linear propagation equation (\ref{psidot}) for appropriate behaviours of 
$a(t).$

\subsection{An Approximation Method}

We consider spatially flat universes ($k=0$) and assume that the expansion
scale factor is that of the Friedmann model containing a perfect fluid: 
\begin{equation}
a=t^n  \label{a}
\end{equation}
where $n$ is a constant. The late stages of an open universe containing
fluid with density $\rho $ and pressure $p$ obeying $\rho +3p>0$ can be
studied by considering the case $n=1.$ We rewrite the wave equation (\ref
{psidot}) as 
\[
\ (\dot{\psi}a^3\dot{)}=-\frac{2\zeta }\omega \rho _ma^3\exp [-2\psi ] 
\]
Therefore, since $\rho _ma^3$ is constant this reduces to a Liouville
equation of the form

\begin{equation}
(\dot{\psi}a^{3})^{.}=\ N\exp [-2\psi ]  \label{psi}
\end{equation}
where $N$ is a constant, defined by 
\begin{equation}
N\equiv -\frac{2\zeta }{\omega }\rho _{m}a^{3}\ .  \label{N}
\end{equation}
We shall consider first the cosmological models that arise when the defining
constant $N$ is negative. This arises when the constant $\zeta <0,$
indicating that the matter content of the universe is dominated by magnetic
rather than electrostatic energy. The value of $\zeta $ for baryonic and
dark matter has been disputed \cite{zal,olive,bsbm}. It is the difference
between the percentage of mass in electrostatic and magnetostatic forms. As
explained in \cite{bsbm}, we can at most \textit{estimate} this quantity for
neutrons and protons, with $\zeta _{n}\approx \zeta _{p}\sim 10^{-4}$. We
may expect that for baryonic matter $\zeta \sim 10^{-4}$, with
composition-dependent variations of the same order. The value of $\zeta $
for the dark matter, for all we know, could be anything between -1 and 1.
Superconducting cosmic strings, or magnetic monopoles, display a \textit{%
negative} $\zeta $, unlike more conventional dark matter. It is clear that
the only way to obtain a cosmologically increasing $\alpha $ in BSBM is with 
$\zeta <0$, i.e with unusual dark matter, in which magnetic energy dominates
over electrostatic energy. In \cite{bsbm} we showed that fitting the Webb et
al results requires $\zeta /\omega =-2\pm 1\times 10^{-4}$, where $\zeta $
is weighted by the necessary fractions of dark and baryonic matter required
by observations of the gravitational effects of dark matter and the
calculations of Big Bang nucleosynthesis. We note also that in practice $%
\zeta $ might display a significant spatial variation because of the change
in the nature of the dominant form of dark matter over different length
scales. For example, a magnetically dominated form of dark matter might
contribute a negative value of $\zeta $ on large scales while domination of
the matter content by baryons on small scales would lead to $\zeta >0$
locally. We will not discuss the effects of such variations in this paper.

\subsection{The Validity of the Approximation}

\bigskip We have assumed that the scale factor is given by the FRW model and
then solved the $\psi $ evolution equation. This is a good approximation up
to logarithmic corrections. Here is what happens to higher order.

We take the leading order behaviour in (\ref{fried})

\begin{equation}
\left( \frac{\dot{a}}{a}\right) ^{2}\approx \frac{8\pi }{3}\rho _{m}\left(
1+\left| \zeta \right| e^{-2\psi }\right)  \label{next}
\end{equation}
Now if we take $a=t^{2/3}$ so

\[
\frac{8\pi }{3}\rho _{m}=\frac{4}{9t^{2}} 
\]
and solve equation (\ref{psi}) we get asymptotically,

\begin{equation}
\psi =\frac{1}{2}\ln [2N\ln t]\rightarrow \ln [\ln t]  \label{psisol}
\end{equation}
to leading order at late times. Suppose we now re-solve (\ref{next}) with
the $\left| \zeta \right| e^{-2\psi }$ correction included

\begin{equation}
\left( \frac{\dot{a}}{a}\right) ^{2}\approx \frac{4}{9t^{2}}\ \left(
1+\left| \zeta \right| e^{-2\psi }\right) \approx \frac{4}{9t^{2}}\left( 1+%
\frac{\left| \zeta \right| }{\ \ln t}\right)  \label{newfried}
\end{equation}
Note that the kinetic term which we neglected is of order

\begin{equation}
\omega \dot{\psi}^{2}\sim \frac{1}{4t^{2}\ln ^{2}t}
\end{equation}
and so is smaller than the term we have retained. Solving equation (\ref
{next}) we have

\begin{equation}
\ln a=\frac{2\left| \zeta \right| }{3}\left[ sh^{-1}\left\{ \sqrt{\frac{\ln t%
}{\left| \zeta \right| }}\right\} +\sqrt{\frac{\ln t}{\left| \zeta \right| }}%
\sqrt{1+\frac{\ln t}{\left| \zeta \right| }}\right]
\end{equation}
Note that when $\left| \zeta \right| \rightarrow 0$ this gives the usual $%
\ln a=\frac{2}{3}\ln t$. When $\left| \zeta \right| \neq 0$ we have

\[
\ln a\rightarrow \frac{2\left| \zeta \right| }{3}\left\{ \ln [2\sqrt{\frac{%
\ln t}{\left| \zeta \right| }}]+\frac{\ln t}{\left| \zeta \right| }\right\} 
\]
and

\begin{equation}
a=t^{2/3}(\ln t)^{\left| \zeta \right| /3}  \label{scale}
\end{equation}
where $\left| \zeta \right| $ is small and so the corrections to the $%
a=t^{2/3}$ ansatz are small. In terms of the Hubble rate:

\[
H=\frac{2}{3t}+\frac{\left| \zeta \right| }{3t\ln t} 
\]

If we include the kinetic corrections to equation (\ref{next}) then as $\dot{%
\psi}=\frac{1}{2t\ln t}$

\begin{equation}
\left( \frac{\dot{a}}{a}\right) ^{2}\approx \frac{4}{9t^{2}}\ \left(
1+\left| \zeta \right| e^{-2\psi }\ \right) +\frac{4\pi \omega }{3}\dot{\psi}%
^{2}\approx \frac{4}{9t^{2}}\left( 1+\frac{\left| \zeta \right| }{\ \ln t}+%
\frac{S}{\ \ln ^{2}t}\right)
\end{equation}
where

\[
S\equiv \frac{3\pi \omega }{4} 
\]

So, if $x=\ln t,$we have

\begin{eqnarray}
\frac{3}{2}\ln a &=&\sqrt{x^{2}+\left| \zeta \right| x+S}+S\int \frac{dx}{x%
\sqrt{x^{2}+\left| \zeta \right| x+S}}  \nonumber \\
&&+\frac{\left| \zeta \right| }{2}\int \frac{dx}{\sqrt{x^{2}+\left| \zeta
\right| x+S}} \\
\frac{3}{2}\ln a &=&\sqrt{x^{2}+\left| \zeta \right| x+S}-\ln \left[ \frac{%
2S+\left| \zeta \right| x+2\sqrt{S}\sqrt{x^{2}+\left| \zeta \right| x+S}}{x}%
\right]  \nonumber \\
&&+\frac{\left| \zeta \right| }{2}\ln \left[ 2\sqrt{x^{2}+\left| \zeta
\right| x+S}+2x+\left| \zeta \right| \right]
\end{eqnarray}
Again, as $t\rightarrow \infty $ the leading order behaviour is that found
in equation (\ref{scale}).

In the radiation era we have an exact solution of equation (\ref{psi}) with

\begin{equation}
2\psi =\ln (8N)+\frac{1}{2}\ln t
\end{equation}
so the corrections to the Friedmann equation look like

\begin{equation}
\left( \frac{\dot{a}}{a}\right) ^{2}\approx \frac{1}{4t^{2}}\ \left(
1+\left| \zeta \right| e^{-2\psi }\right) \approx \frac{1}{4t^{2}}\left( 1+%
\frac{\left| \zeta \right| }{8N\ t^{1/2}}\right)
\end{equation}
and these corrections fall off much faster than in the dust case. Again, our
basic approximation method holds good to high accuracy.

\subsection{A Linearisation Instability}

Despite the robustness of the basic test-motion approximation that we are
employing to analyse the evolution of $\psi (t)$ as the universe expands,
there is a subtle feature the non-linear evolution equation (\ref{psi})
which must be noted in order that spurious conclusions are not drawn from an
approximate analysis. We see that the right-hand side of equation (\ref{psi}%
) is always positive. Therefore $\psi $ can never experience an expansion
maximum (where $\dot{\psi}=0$ and $\ddot{\psi}<0$) and therefore $\psi (t)$
can never oscillate. However, if we were to linearise equation (\ref{psi}),
obtaining

\[
(\dot{\psi}a^{3})^{.}=\ N\exp [-2\psi ]\approx N(1-2\psi +O(\psi ^{2})) 
\]
then for $\psi >1/2$ the right-hand side takes negative values and
pseudo-oscillatory solutions for $\psi $ would appear that are not the
linearised approximation to any true solution of the non-linear equation (%
\ref{psi}). Care must therefore be taken to ensure that analytic
approximations are not extended to large $\psi $ and that numerical analyses
are not creating spurious spirals in the phase plane by virtue of a
linearisation procedure; for a fuller discussion see ref. \cite{bsm3}.

These considerations can be taken further. It is possible for $\psi (t)$ to
decrease, reach a minimum and then increase. But it is not possible for $%
\psi (t)$ to decrease if it has ever increased. A second interesting
consequence of this feature of equation (\ref{psi}) is that it holds true
even if $a(t)$ reaches an expansion maximum and begins to contract. Thus in
a closed universe we expect $\psi $ and $\alpha $ to continue to increase
slowly even after the universe begins to contract. This will have important
consequences for the expected variation of $\psi $ and $\alpha $ in
realistically inhomogeneous universes.

\section{Phase-plane Analysis}

\subsection{A transformation of variables}

In this section we will look at the $\psi $ equation of motion (\ref{psi}%
)when the expansion scale factor takes the power-law form (\ref{a}). The
evolution equation for the field then becomes:

\begin{equation}
\frac d{dt}\left( \dot{\psi}t^{3n}\right) =N\exp [-2\psi ]  \label{n}
\end{equation}
with $N>0.$ We introduce the following variables :

\begin{eqnarray}
x=ln(t)\qquad y=\psi -Ax-B\qquad A=1-\frac{3n}2\qquad B=\frac 12\ln (N)
\label{coord}
\end{eqnarray}
and rewrite (\ref{psi}) as:

\begin{equation}
y^{\prime \prime }+(3n-1)(y^{\prime }+1-\frac{3n}2)=e^{-2y},  \label{y}
\end{equation}
where $^{\prime }$ $\equiv d/dx$. This second-order differential equation
can be transformed into an autonomous system by defining $v=\frac{dy}{dx}$
and $u=y$:

\begin{eqnarray}
\frac{dv}{dx} &=&e^{-2u}+(1-3n)(v+1-\frac{3n}2)  \label{nl} \\
\frac{du}{dx} &=&v  \nonumber
\end{eqnarray}

We see that the system has a finite critical point at

\begin{equation}
(u_c,v_c)=(-\frac 12\ln {[(1-3n)(\frac{3n}2-1)]},0)  \label{crit}
\end{equation}

when $n\in (\frac 13,\frac 23)$ and it has a infinite critical point at $%
(u_c,v_c)=(+\infty ,0)$ when $n=\frac 13$ or $n=\frac 23$. The finite
critical points correspond to the family of exact solutions of the form $%
\psi =B+A\ln (t)$ found in \cite{bsm1}. In the original variables these
solutions are

\[
\psi =\frac 12\ln [\frac{2N}{(2-3n)(3n-1)}]+\left( \frac{2-3n}2\right) \ln
(t). 
\]

In order to analyse the system fully, we will study finite critical points
and the critical points at infinity separately. We will also distinguish
several domains of behaviour for $n$: $n=\frac 13$, $n=\frac 23$, $n\in
(\frac 13,\frac 23)$, $n<\frac 13$ and $n>\frac 23$. In some cases we will
supplement our investigations by numerical study of the system (\ref{nl}).

\subsection{The Finite Critical Points}

\subsubsection{The Cases $n\in (\frac 13,\frac 23)$}

In the domain where $n\in (\frac 13,\frac 23)$ the system (\ref{nl}) does
not appear to be exactly integratable. So in this section we will study the
behaviour of the system near the critical points and at late times.

There is a finite critical point $(u_c,v_c)=(-\frac 12\ln {[(1-3n)(\frac{3n}%
2-1)]},0)$ for $n\in (\frac 13,\frac 23)$. Linearising the system \ref{nl}
about it we obtain:

\begin{eqnarray}
\frac{dV}{dx} &=&(3n-1)(3n-2)U+(1-3n)V  \label{almost} \\
\frac{dU}{dx} &=&V  \nonumber
\end{eqnarray}
where $V=v-v_c$ and $U=u-u_c$. Since the characteristic matrix is non
singular the critical point is simple. Hence the non-linearised version of
this system has the same phase portrait at the neighbourhood of the critical
point.

The eigenvalues $\xi _{1,2}$ and corresponding eigenvectors $\chi ^{\xi
_{1,2}}$ of the system (\ref{almost}) are:

\begin{eqnarray}
\xi _{1,2} &=&{\frac{1-3n\pm \sqrt{9-42n+45{n^2}}}2} \\
\chi ^{\xi _{1,2}} &=&({\frac{1-3n\pm \sqrt{9-42n+45{n^2}}}2},1)  \nonumber
\end{eqnarray}
These eigenvalues will be complex numbers for $n\in (\frac 13,\frac 35)$ and
they will be pure real numbers when $n\in (\frac 35,\frac 23)$. Notice that
for both cases the real part of the eigenvalues is always negative, so the
critical point is a stable attractor. The general solution of the linearised
system (\ref{almost}) can be expressed as:

\[
(U(x),V(x))=A{e^{\xi _1x}}\chi _1+B{e^{\xi _2x}}\chi _2 
\]
where $A$, $B$ are arbitrary constants and $x=\ln (t)$. From the
transformations (\ref{psi}) we can obtain explicitly the expression for $%
\psi (t)$ near the critical point. There are three possible behaviours of
the solutions near the critical point:

\subsubsection{Pseudo-Oscillatory $\protect\psi $ behaviour}

When $n\in (\frac{1}{3},\frac{3}{5})$ the linearised evolution for the $\psi 
$ field exhibits damped oscillations about its asymptotic solution as $%
t\rightarrow \infty $:

\begin{eqnarray}
\psi (t) &=&\frac 12\ln [\frac{2N}{\left( 1-3n\right) \left( 3n-2\right) }%
]+\frac 12\left( 2-3n\right) \ln (t)  \label{comp} \\
&&+{t^\alpha }\left( B\cos [\beta \ln (t)]+A\sin [\beta \ln (t)]\right) 
\nonumber
\end{eqnarray}
where

\begin{eqnarray}
\alpha  &=&{\frac{1-3n}{2}}  \label{alph} \\
\beta  &=&{\frac{\sqrt{9+42n-45{n^{2}}}}{2}}  \label{beta}
\end{eqnarray}
We note that this family of solutions and asymptotes includes the case of
the radiation-dominated universe ($n=1/2$). The phase portrait and the $\psi 
$ vs. $\ln t$ for this case are shown in Figure 1. 

\begin{figure}[ht]
\centering
\epsfig{file=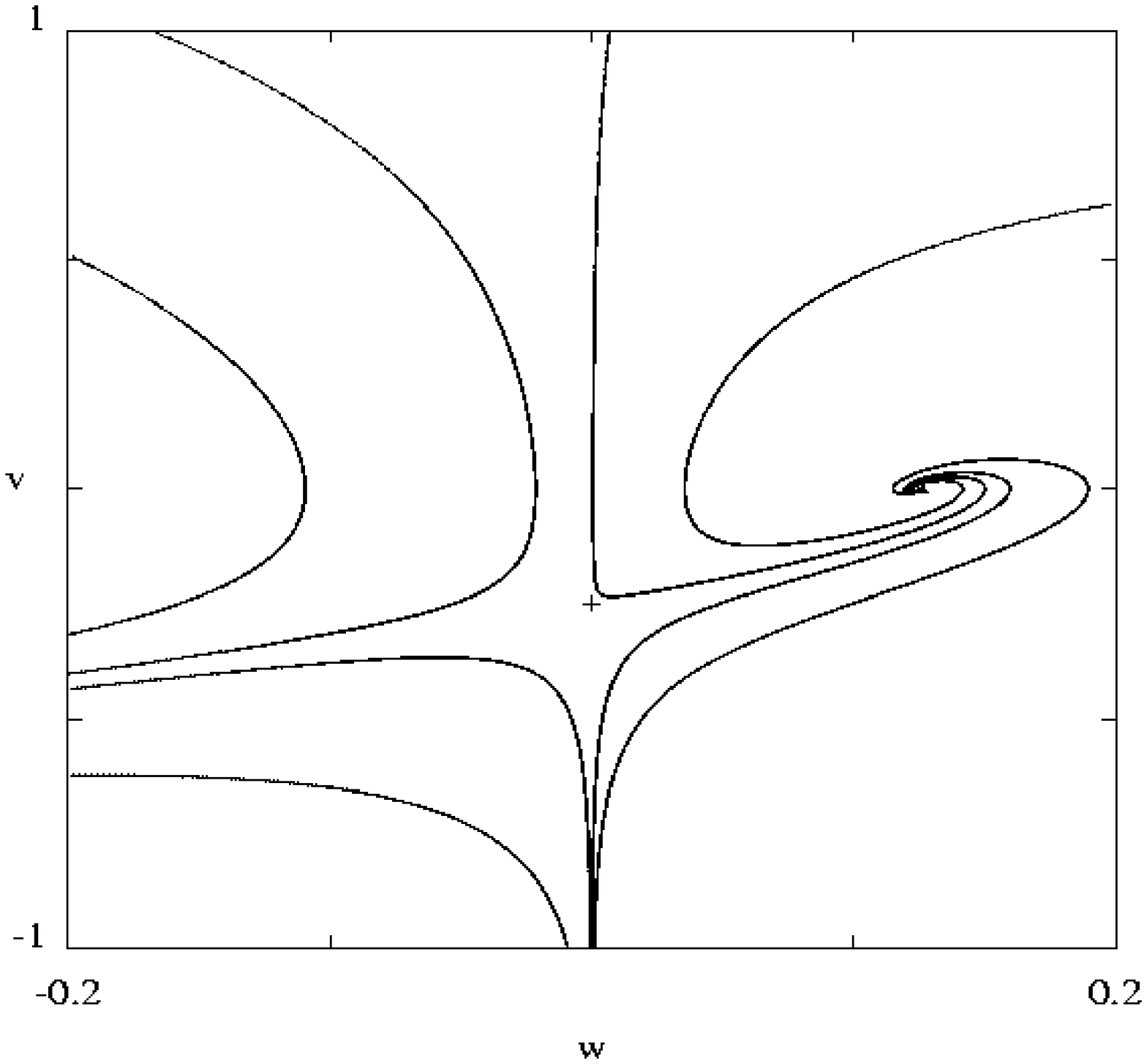,height=6cm} %
\epsfig{file=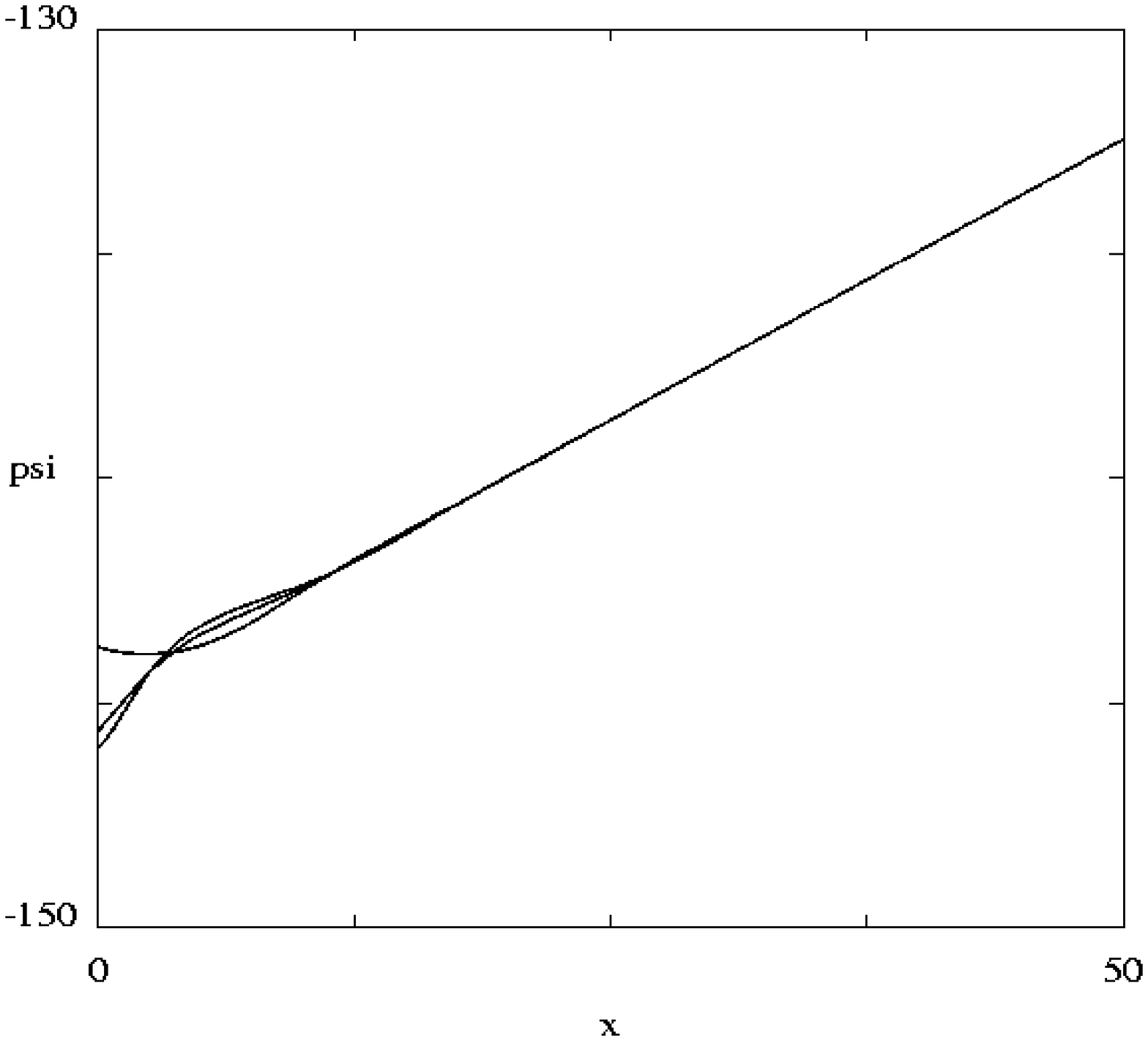,height=6cm}
\caption{{\protect\small \textit{Numerical plots of the phase space in the $%
(w,v)$ coordinates, and the $\protect\psi$ evolution with $x=log(t)$ for $n=%
\frac{1}{2}$ .The '$+$' sign is a saddle point and the triangle is a stable
node. }}}
\label{n12}
\end{figure}

But in the case of a universe containing the balance of matter and
radiation, displayed by our own, it need not be the case that the asymptotic
behaviour, displayed by the exact solution for the critical point, is
reached before the radiation-dominated expansion is replaced by
matter-domination, see references \cite{bsbm} and \cite{bsm1} for further
discussion of this point. However as we discussed above these oscillations
are an artefact of the linearisation process and the part of the solution (%
\ref{comp}) controlled by the constants $A$ and $B$ is only valid for small
times, hence we called this behaviour pseudo-oscillatory.

\subsubsection{Non-oscillatory behaviour}

When $n\in (\frac 35,\frac 23)$, the $\psi $ field will approach its
asymptotic behaviour in a non-oscillatory fashion as $t\rightarrow \infty $:

\begin{eqnarray}
\psi (t) &=&\frac 12\ln [\frac{2N}{\left( 1-3n\right) \left( 3n-2\right) }]
\label{real} \\
&&+\frac 12\left( 2-3n\right) \ln (t)+t^{\delta -\gamma }\left( A+B{%
t^{2\gamma }}\right)  \nonumber
\end{eqnarray}
where

\begin{eqnarray}
\delta &=&{\frac{1-3n}2} \label{del} \\ 
\gamma &=& \frac 12\sqrt{9-42n+45{n^2}}  \label{gam}
\end{eqnarray}

Note that $\delta +\gamma <0$ in the domain $n\in (\frac{1}{3},\frac{2}{3})$%
, so at late times ($t\rightarrow \infty $), both the pseudo-oscillatory and
non-oscillatory solutions case will approach the asymptotic solution defined
by the appropriate value of $n$.

\subsubsection{Intermediate behaviour}

A transition between the pseudo-oscillatory and non-oscillatory regimes
happens when $\beta =0$ and $n=\frac{3}{5}$; the $\psi $ field then has the
following solution in the vicinity of the critical point:

\begin{equation}
\psi (t)=\frac{1}{5}\ln (N)+\frac{1}{10}\ln (t)+At^{-{\frac{2}{5}}}.
\label{n3/5}
\end{equation}
The phase plane structure and evolution of $\psi $ vs. $\ln t$ for this case
are shown in Figure 2.

\begin{figure}[ht]
\centering
\epsfig{file=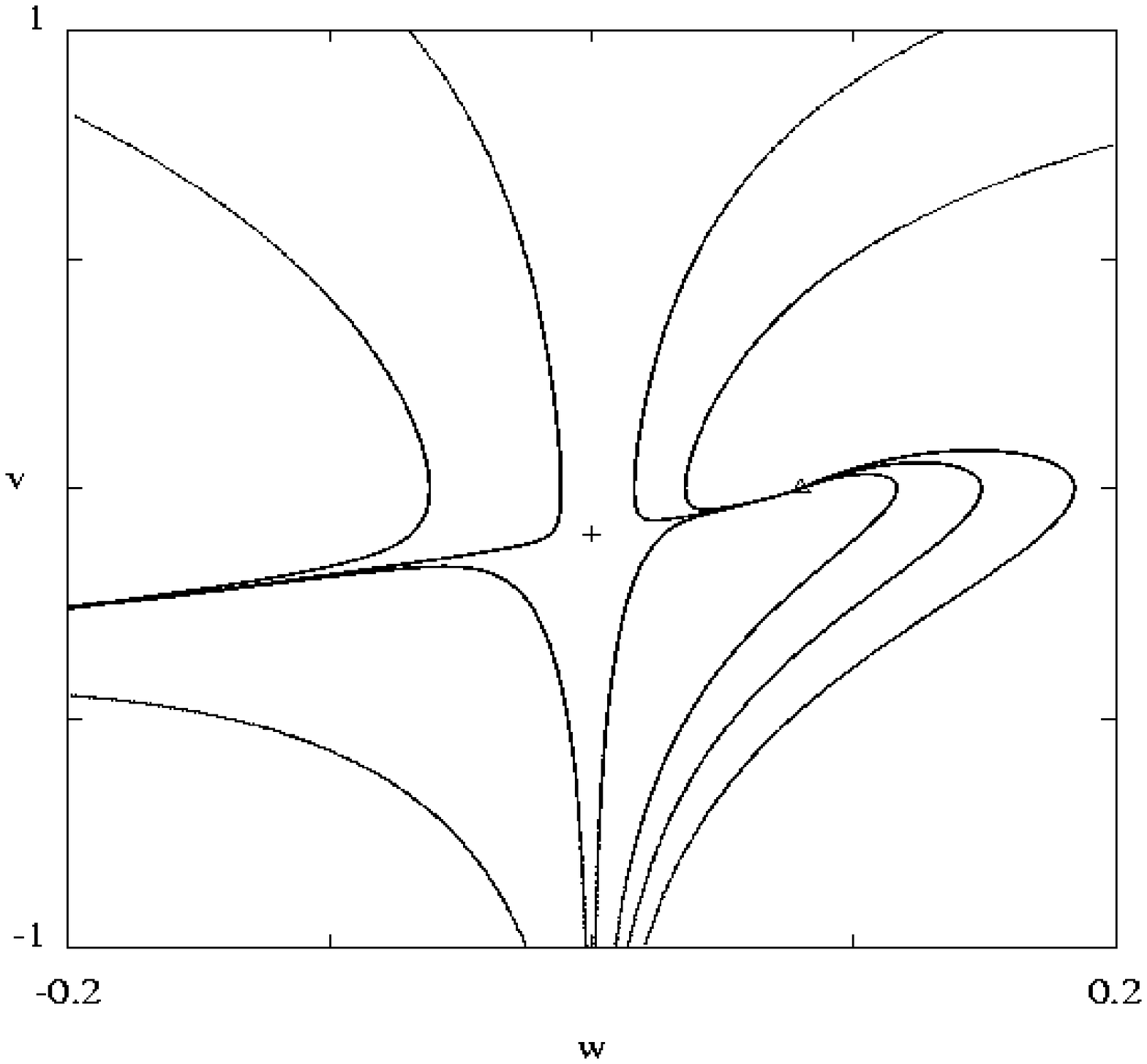,height=6cm} %
\epsfig{file=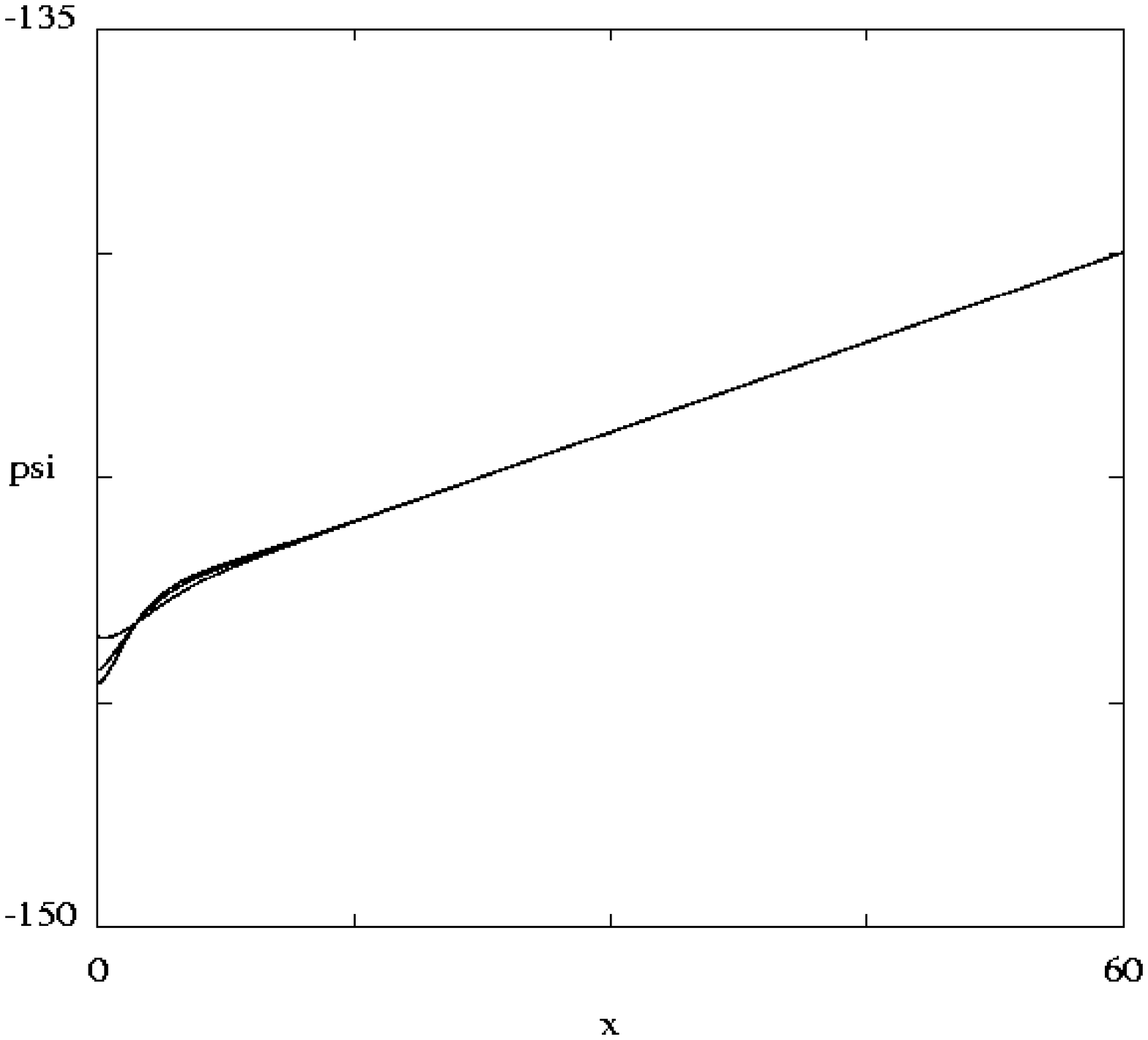,height=6cm}
\caption{{\protect\small \textit{Numerical plots of the phase space in the $%
(w,v)$ coordinates, and the $\protect\psi$ evolution with $x=log(t)$ for $n=%
\frac{3}{5}$. The '$+$' sign is a saddle point and the triangle is a stable
node. }}}
\label{n35}
\end{figure}

\subsubsection{Overview}

The late-time solution of the $\psi $ field solution is given by asymptotic
behaviour of equations (\ref{comp}) or (\ref{real}), which is:

\[
\psi \rightarrow \frac 12\ln [\frac{2N}{\left( 1-3n\right) \left(
3n-2\right) }]+\frac 12\left( 2-3n\right) \ln (t) 
\]
This shows that the solutions (\ref{comp}) and (\ref{real}) generalise the
ones found in \cite{bsm1}, since they can be obtained setting $A=B=0$ in (%
\ref{comp}). In particular, the case of a radiation-dominated Universe ($%
n=\frac 12$), we have from (\ref{comp}):

\begin{eqnarray}
\psi (t) &=&\frac{1}{2}\left( \log (8)+\log (N)+\log (t)\right)  \label{rad}
\\
&&+{t^{{-\frac{1}{4}}}}\left( B\cos [{\frac{5\sqrt{3}\log (t)}{4}}]+A\sin [{%
\frac{5\sqrt{3}\log (t)}{4}}]\right)  \nonumber
\end{eqnarray}
A full mathematical summary of the change in structure of the phase space
with changing $n$ for the system (\ref{psi}) is given in the Appendix. There
we include cosmologically unphysical values of $n$ and show how the critical
point structure bifurcates with the change in value of $n$.

\subsection{The Critical Points at Infinity}

In order to describe the qualitative evolution of the system, we must
determine the behaviour of the system (\ref{nl}) near the critical point $%
(u_c,v_c)=(+\infty ,0)$. In order to bring the critical point to a finite
value, we define:

\[
u=-\frac 12\ln (w) 
\]
Using the new coordinate $w$ we can re-write the system (\ref{nl}) as:

\begin{eqnarray}
\frac{dw}{dx} &=&-2wv  \label{nllog} \\
\frac{dv}{dx} &=&\left( 1-3n\right) \left( 1-{\frac{3n}2}+v\right) +w 
\nonumber
\end{eqnarray}

This system has critical points on the $(w,v)$ plane when $(0,-1 + {\frac{3n%
}{2}})$ and $(\left(1- 3n\right) \left( {\frac{3n}2}-1\right),0)$. Note that
the second critical point is just the same as the one we have analysed in
the previous section. In this subsection we will then only analyse the
critical point $(w_c,v_c)=(0,-1 + {\frac{3n}{2}})$ since it corresponds to
the case where $u \rightarrow +\infty$.

Proceeding as before, and linearising (\ref{nllog}) about $%
(w_{c},v_{c})=(0,-1+{\frac{3n}{2}})$ we obtain:

\begin{eqnarray}
\frac{dW}{dx} &=&(2-3n)W  \label{llog} \\
\frac{dV}{dx} &=&W+(1-3n)V  \nonumber
\end{eqnarray}
where $V=v-v_{c}$ and $W=w-w_{c}$. Again, the characteristic matrix of the
system is non singular, and the critical point is simple. Hence, system (\ref
{llog}) will have the same phase portrait as (\ref{nllog}) in the
neighbourhood of the critical point.

The eigenvalues $\xi _{1,2}$ and corresponding eigenvectors $\chi ^{\xi
_{1,2}}$ of the system (\ref{llog}) are:

\begin{eqnarray}
\xi _1 &=&1-3n,\hspace{1.0in}\chi ^{\xi _1}=(0,1)  \label{inf1} \\
\qquad \xi _2 &=&2-3n,\hspace{1.0in}\chi ^{\xi _2}=(1,1)  \label{inf2}
\end{eqnarray}
These eigenvalues are always real. The critical point is an attractive node
for $n>\frac 23$, a saddle point when $n\in (\frac 13,\frac 23)$, and it
will be an unstable point when $n<\frac 13$. The general solution of the
system (\ref{nllog}) in the neighbourhood of the critical point $%
(w_c,v_c)=(0,-1+{\frac{3n}2})$, can be expressed as:

\[
(w(x)-w_c,v(x)-v_c)=A{e^{\xi _1x}}\chi _1+B{e^{\xi _2x}}\chi _2 
\]
where $A$, $B$ are arbitrary constants and $x=\ln (t)$. Therefore, near the
critical point $(u_c,v_c)=(+\infty ,0)$:

\begin{equation}
\psi (t)=\frac 12\ln (\frac NB)  \label{psisolution}
\end{equation}
so the scalar field is constant.

Note that in the domain $n\in (\frac{1}{3},\frac{2}{3})$ this is just a
transitory solution since the critical point is a saddle point. In the
domain $n<\frac{1}{3}$ it is an unstable critical point, possibly relevant
as an early-time solution to braneworld cosmologies in the high-density
regime where the Hubble expansion rate of the universe is linearly
proportional to the density, so $n=1/4$ for a braneworld containing
radiation, $n=1/12$ for a massless scalar field, and $n=1/6$ for dust.
However, in the $t\rightarrow 0$ limit the assumption that the $\dot{\psi}%
^{2}$ and $\zeta \exp [-2\psi ]$ terms can be neglected in the Friedmann
equation (\ref{fried}) will break down.

\subsubsection{The cases of $n>2/3$ and de Sitter expansion}

An interesting case exists when $n>\frac 23$, since the critical point is a
stable attractor and so this means that the constant-$\psi $ behaviour (\ref
{psisolution}) is the late-time attractor, in agreement with the conclusions
of references \cite{bsbm} and \cite{bsm1}. This is an important feature of a
universe which exhibits accelerated expansion in its late stages ($n>1$). It
means that the present value of $\alpha $ is the asymptotic one. It also
means that variations in $\alpha $ are turned off by the domination of the
expansion dynamics by negative curvature or by any quintessence field \cite
{bsbm}, \cite{bsm1}. This property may provide important clues to explaining
why our universe possesses small but finite curvature or quintessence energy
today: if it did not then the fine structure constant would continue to
increase until it was impossible for atoms and molecules to exist \cite{bsm2}%
.

In the $n>2/3$ case we can find a detailed asymptotic solution for equation (%
\ref{n}) which has the form

\[
\psi =C+B(t+t_{0}^{{}})^{1-3n}+\frac{N\exp [-2C]}{2-3n}(t+t_{0}^{{}})^{2-3n}
\]
with $C,t_{0}$ constants. We see immediately that for this solution $\psi $
approaches a constant as $t\rightarrow \infty $. Of particular interest is
the case of a curvature-dominated open universe, which has $n=1$. The phase
plane structure and the evolution of $\psi $ vs. $\ln t$ for this case are
shown in Figure 3.

\begin{figure}[ht]
\centering
\epsfig{file=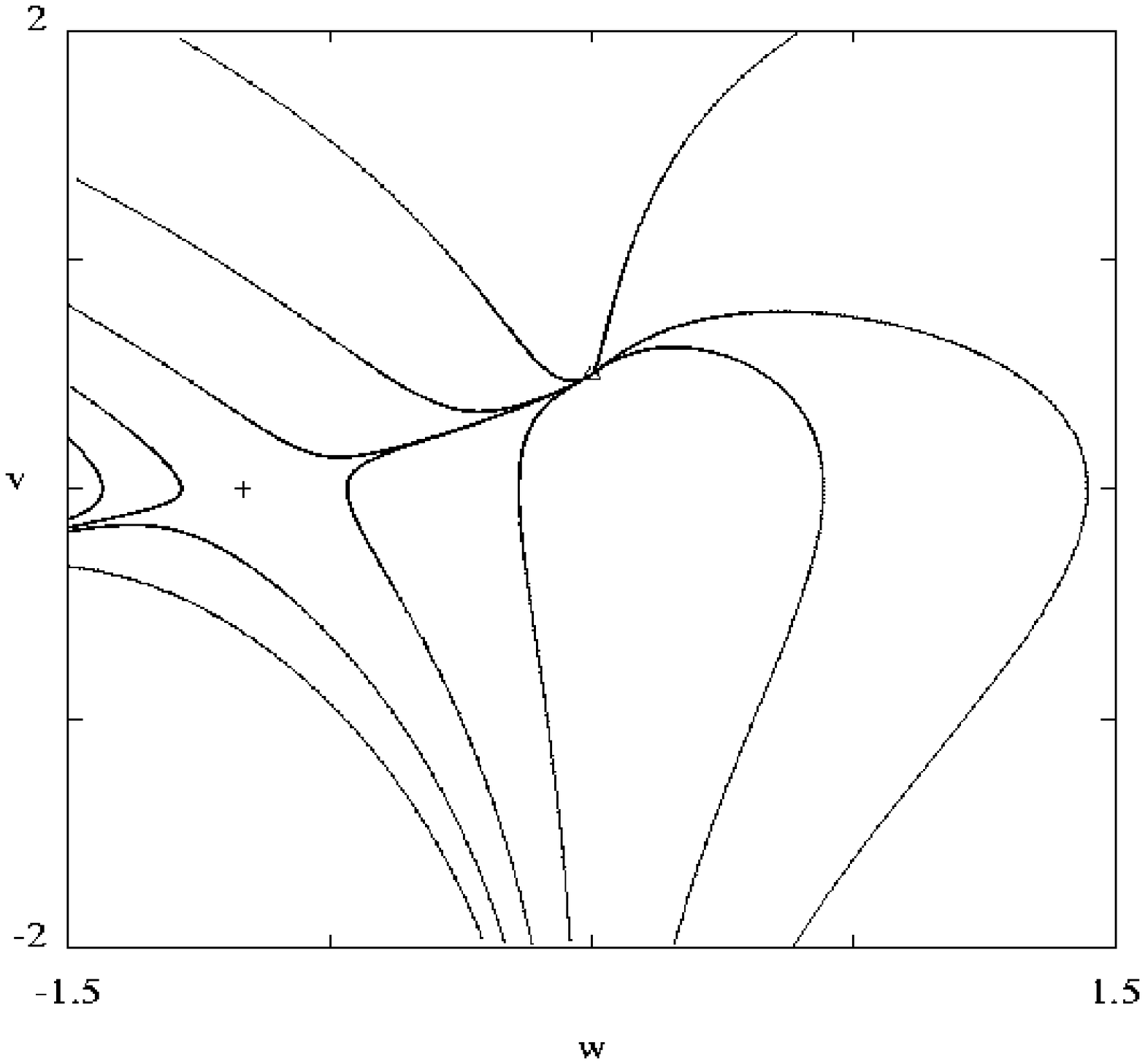,height=6cm} %
\epsfig{file=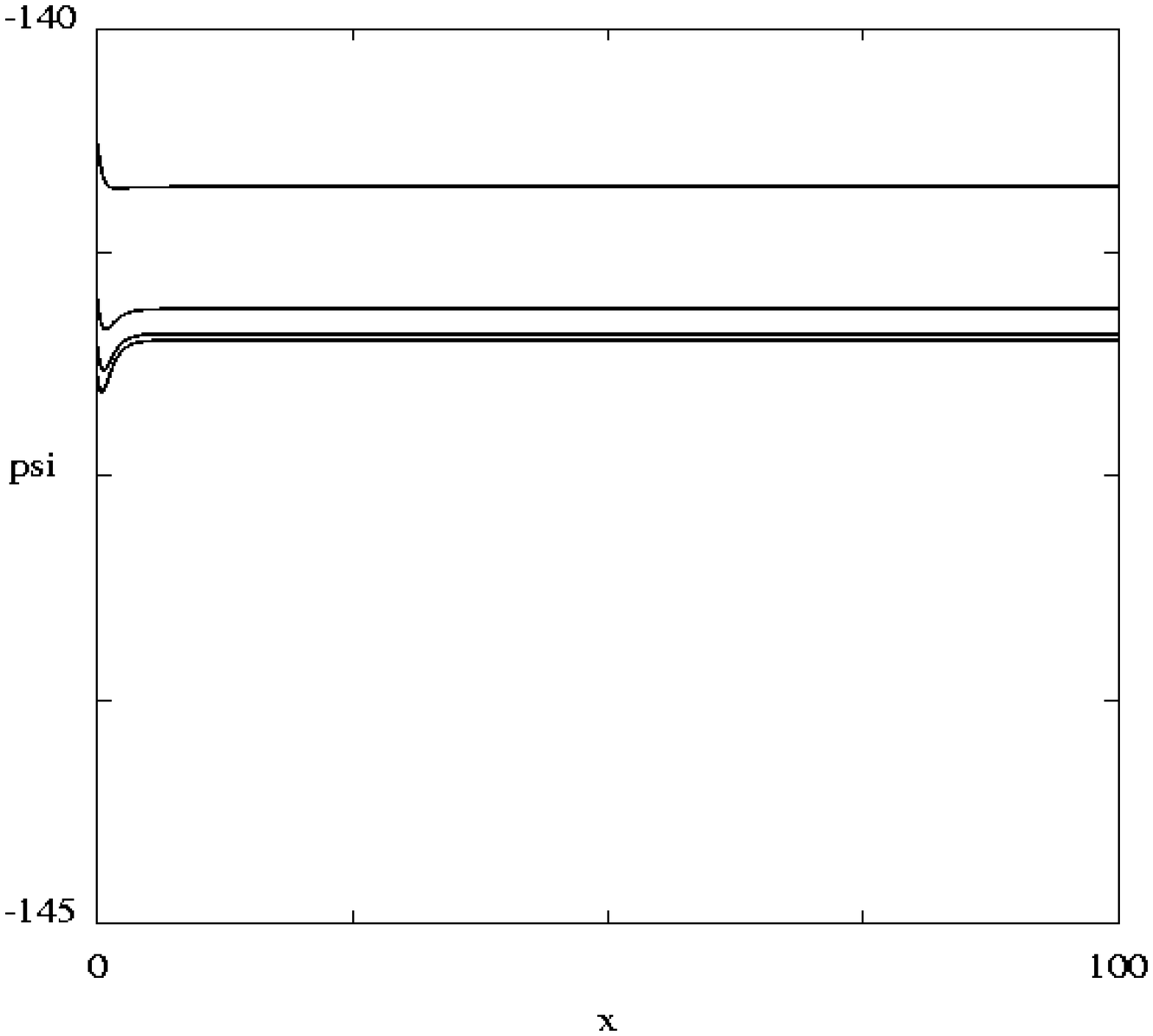,height=6cm}
\caption{{\protect\small \textit{Numerical plots of the phase space in the $%
(w,v)$ coordinates, and the $\protect\psi$ evolution with $x=log(t)$ for $n=1
$. The '$+$' sign is a saddle point and the triangle is a stable node. }}}
\label{n1}
\end{figure}

We see that this approach to constant behaviour occurs for all universes
that accelerate ($n>1$) and so we would expect to find it also in the case
of a de Sitter background universe with

\[
a(t)=\exp [\lambda t]
\]
where $\lambda >0$ is constant. 

Substituting this in equation (\ref{psi}) we find a late-time asymptotic
solution

\[
\psi =C+B\exp [-3\lambda t]-\frac{N\left( 3\lambda t+1\right) }{9\lambda ^2}%
\exp [-2C-3\lambda t]\rightarrow C 
\]
as $t\rightarrow \infty .$

Notice that we were not able to fully describe the nature of this critical
point in the cases where $n=\frac{1}{3}$ or $n=\frac{2}{3}$ since one of the
eigenvalues of the systems is zero. In order to do so, we will study these
two cases individually, in particular the case $n=\frac{2}{3}$ is important
since it describes the scale factor evolution on a dust dominated universe.

\subsubsection{The $n=\frac 13$, $a(t)=t^{\frac 13}$ case: an exact solution}

The $n=\frac 13$ case can be exactly integrated. It is of interest as an
exact solution in its own right but it corresponds to the case of a universe
whose expansion dynamics are dominated by the effects of a fluid with a $%
p=\rho $ equation of state, or a massless scalar field. It also describes
the behaviour of the $\psi $ field in an anisotropic universe of the simple
Kasner type. We see when $n=1/3$ the system (\ref{nl}) has the form:

\begin{eqnarray}  \label{0333system}
\frac{dv}{dx} = e^{-2u} \\
\frac{du}{dx} = v  \nonumber
\end{eqnarray}

This integrates to give

\[
v^2+e^{-2u}=E^2 
\]
where $E$ is a constant. Hence, we have two possible solution branches: $%
v(u)=\pm \sqrt{E-e^{-2u}}$. Both, positive ($+$) and negative ($-$)
solutions for $v$ lead to the same result. Choosing the positive branch of
the $v(u)$ solution and using equations (\ref{0333system}) we obtain:

\[
u(x)=-\ln (E)+\ln [\cosh \left( E\left( C_1-x\right) \right) ] 
\]
where $C_1$ is an integration constant. From (\ref{coord}) we have a
solution for $\psi (t)$:

\[
\psi (t)={\frac{\ln (\frac N4)\ }2}-\ln (E)+\frac 12\ln (t)+\ln [\exp
[EC_1]t^{-E}+t^E\exp [-EC_1]] 
\]
and the asymptotic limit when $t\rightarrow \infty $ gives

\[
\psi (t)\rightarrow (\frac{1}{2}+E)\ln (t+t_{0}).
\]
with $t_{0}$ constant. The phase space and $\psi (t)$ vs. $\ln t$ evolution
is shown in Figure 4.

\begin{figure}[hb]
\centering
\epsfig{file=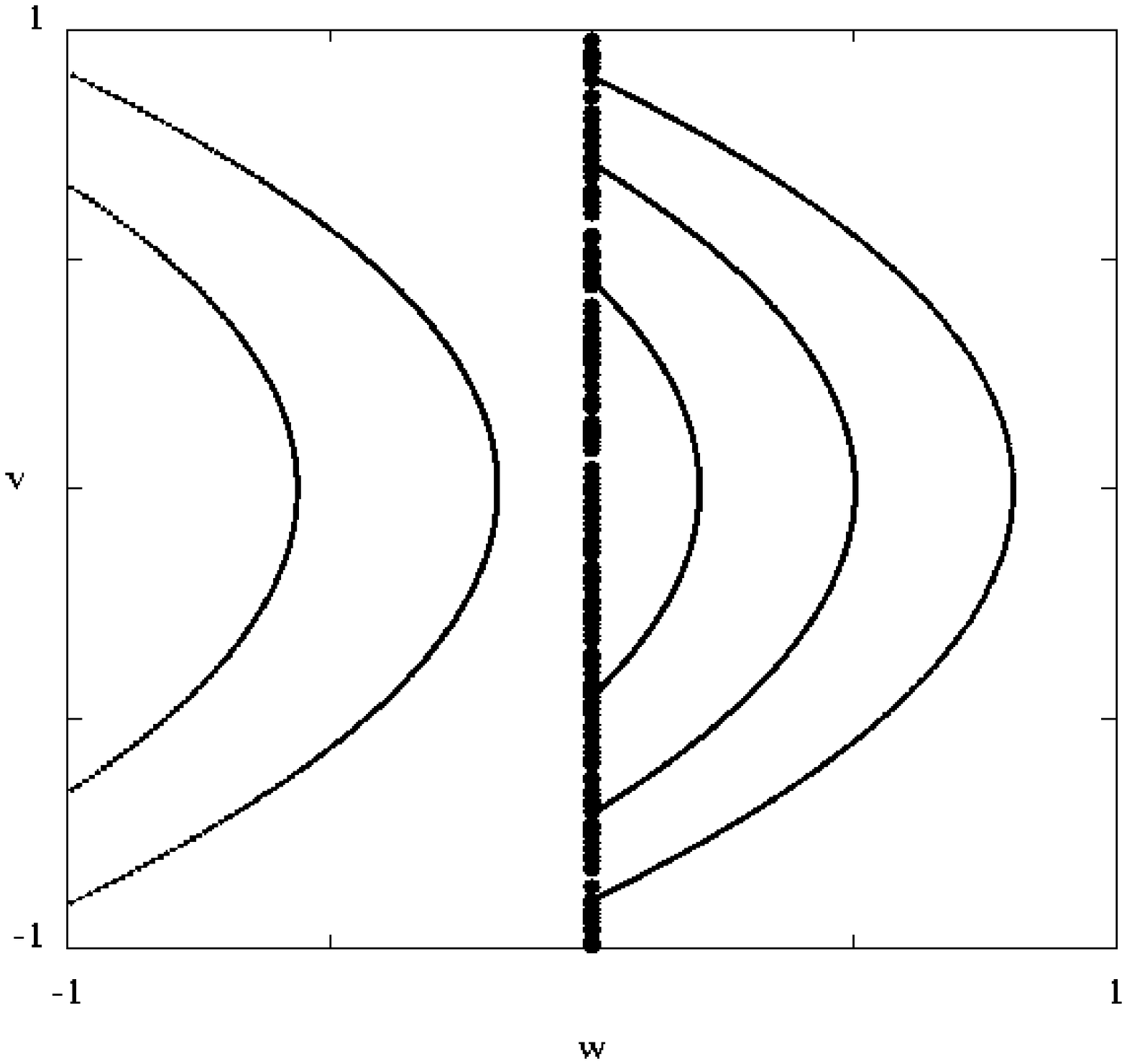,height=6cm} %
\epsfig{file=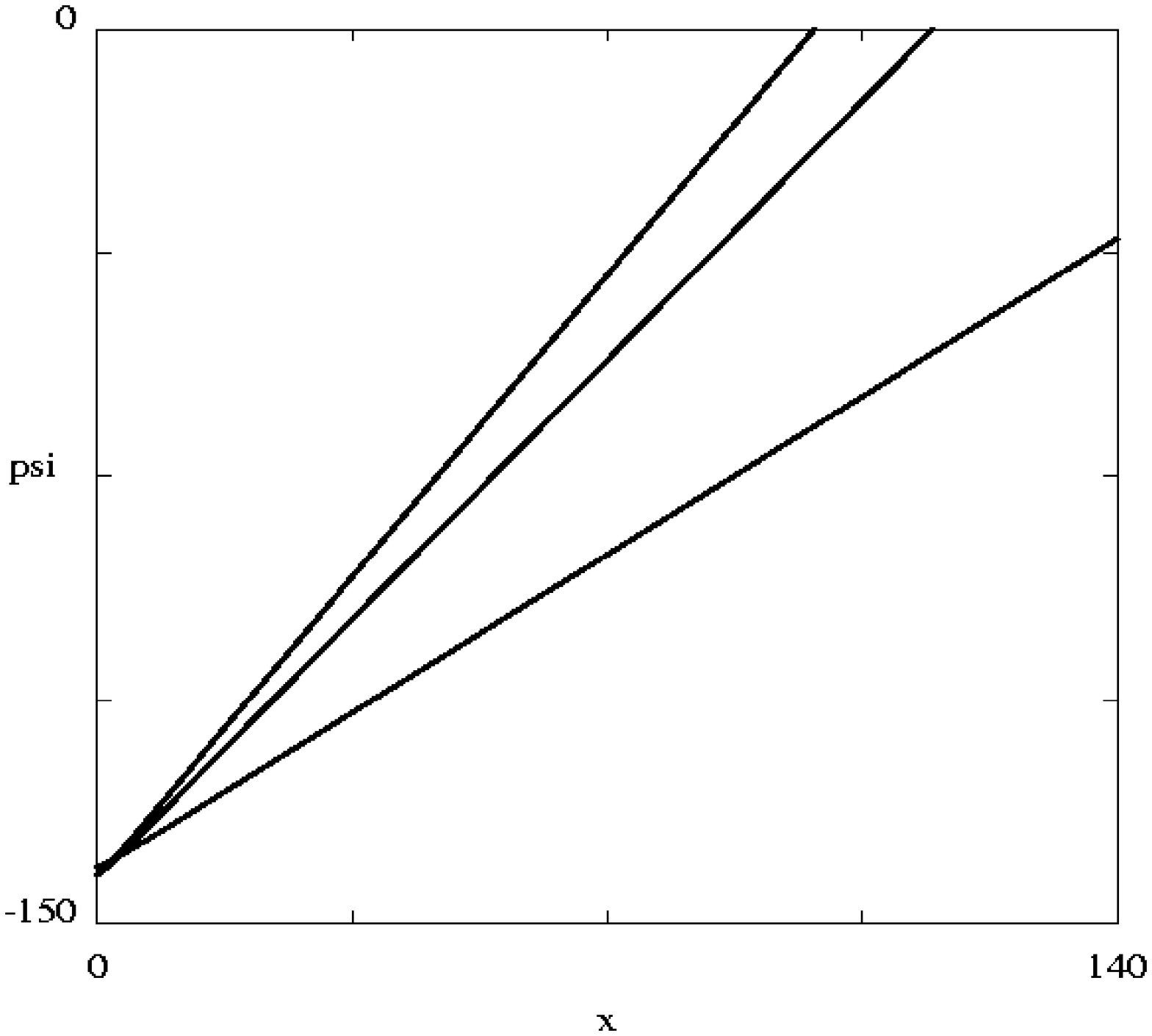,height=6cm}
\caption{{\protect\small \textit{Numerical plots of the phase space in the $%
(w,v)$ coordinates and time evolution of $\protect\psi(x)$ $n=\frac{1}{3}$.
The central line which passes through the origin is an attractor for $w>0$,
and an unstable line for $w<0$. }}}
\label{n13}
\end{figure}

\subsubsection{The $n=\frac 23$ , $a=t^{\frac 23}$ case}

The case of a dust-dominated universe is mathematically special because of
the presence of a zero eigenvalue in the stability analysis. It is
illuminating to consider this case separately with an asymptotic analysis
that extends the earlier study in \cite{bsm1}. Consider equation (\ref{n})
with $n=2/3$. If we introduce the new variable $x=\ln (t),$ then

\begin{equation}
\psi ^{\prime \prime }+\psi ^{\prime }=N\exp [-2\psi ]  \label{dash}
\end{equation}

At large $x$, the asymptotic form of this equation has the form:

\begin{equation}
\psi =\frac{1}{2}\ln [2N(x+x_{0})]-\frac{1}{2}\sum_{n=1}^{\infty }\frac{%
(n-1)!}{(x+x_{0})^{n}}+C\exp [-x]  \label{asym}
\end{equation}
where $C$ and $x_{0}>0$ are arbitrary constants. The leading order behaviour
as $t\rightarrow \infty $ is therefore (cf.(\ref{psisol}))

\[
\psi \sim \frac{1}{2}\ln [2N\ln (t)] 
\]

\paragraph{Stability analysis of the $n=2/3$ asymptote}

\begin{figure}[ht]
\centering
\epsfig{file=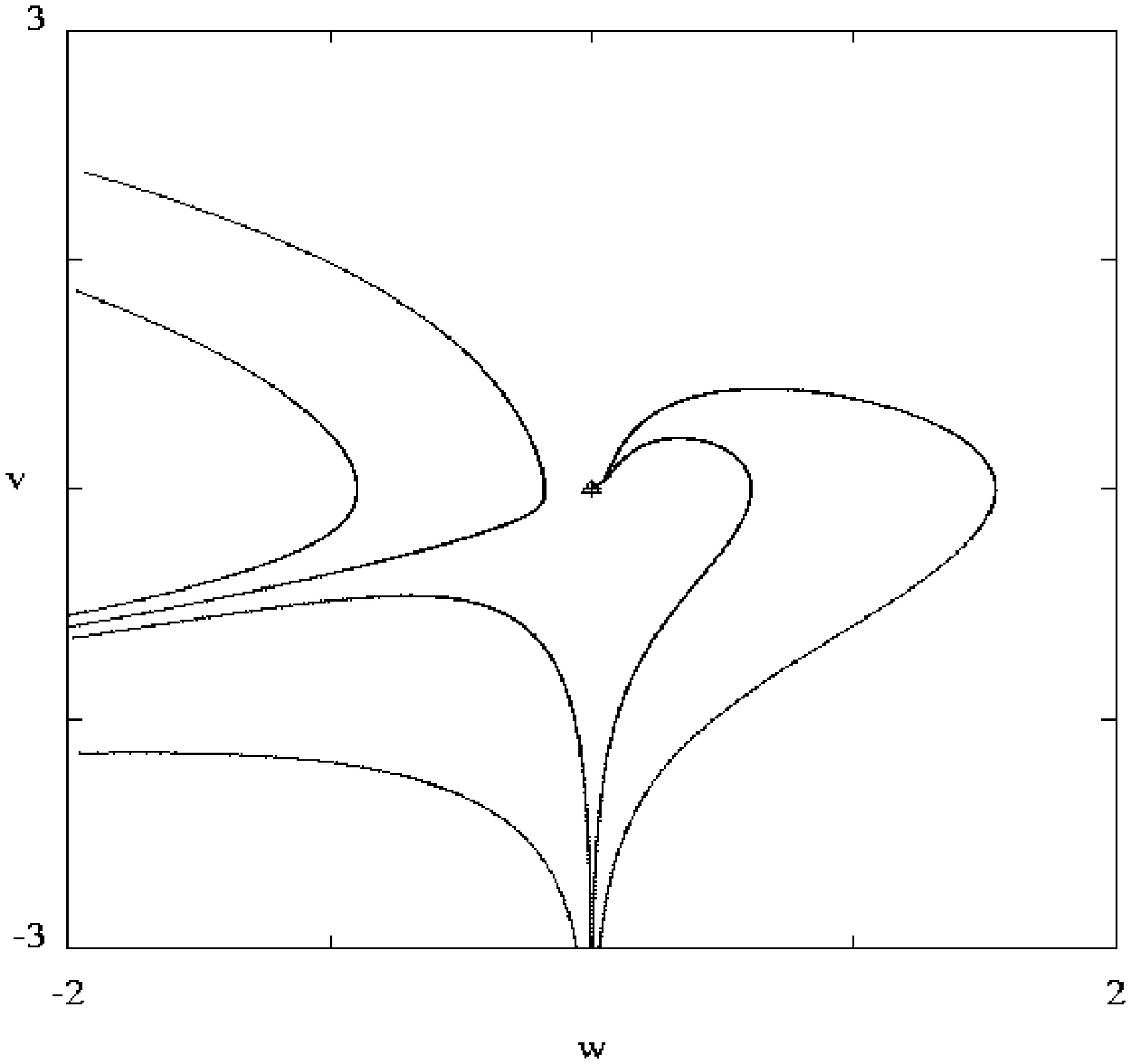,height=6cm} %
\epsfig{file=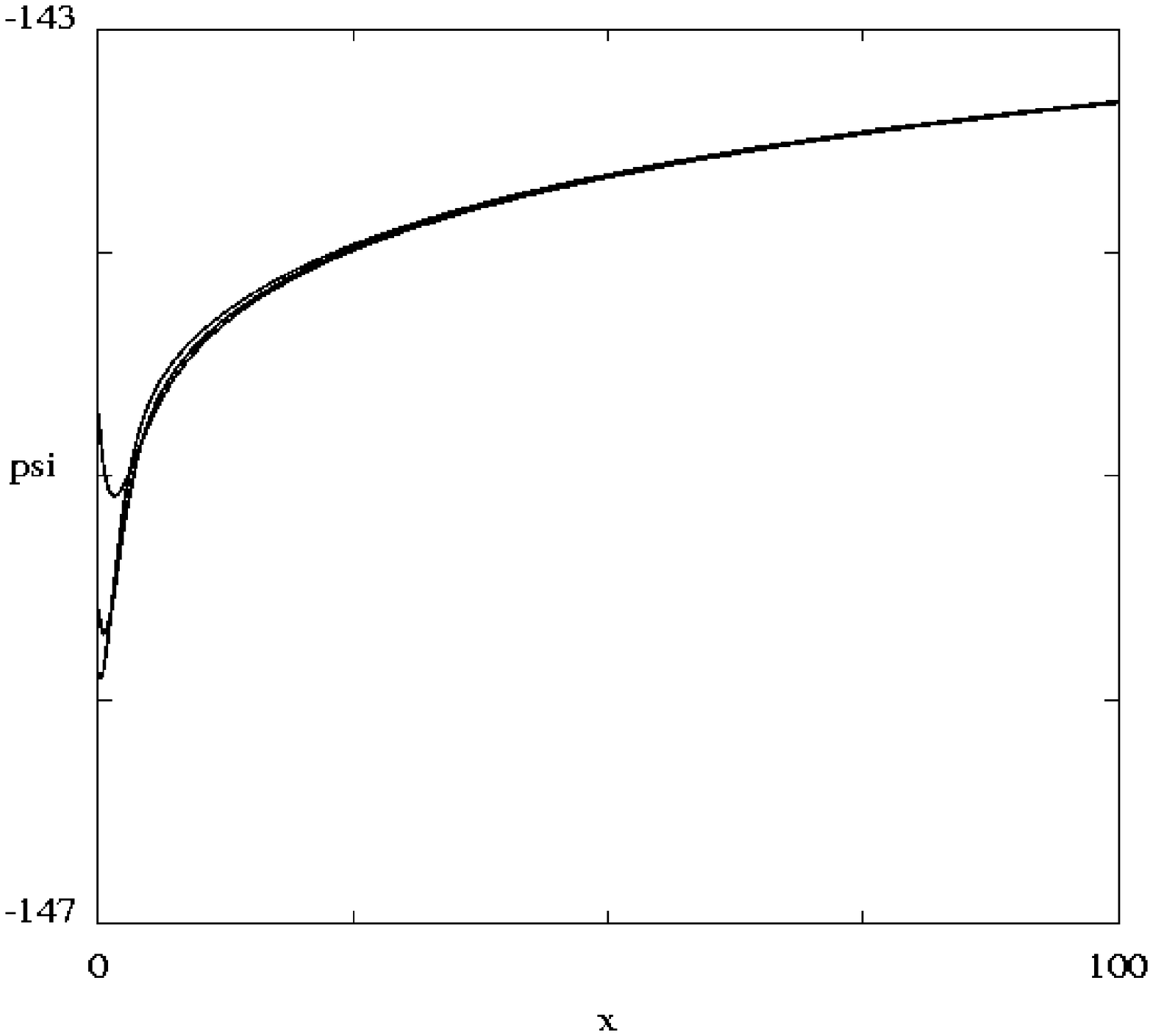,height=6cm}
\caption{{\protect\small \textit{Numerical plots of the phase space in the $%
(w,v)$ coordinates, and the $\protect\psi$ evolution with $x=log(t)$ for $n=%
\frac{2}{3}$. The triangle is a saddle point for $w<0$ and a stable node for 
$w>0$. }}}
\label{n23}
\end{figure}

Performing the coordinate transformation $u=-\frac 12\ln (w)${\ on the
system (\ref{nl}) when }$n=2/3${, we have: }

\begin{eqnarray}
\frac{dv}{dx} &=&w-v  \label{0667non} \\
\frac{dw}{dx} &=&-2vw  \nonumber
\end{eqnarray}
{This system has a critical point at the origin of the }$(w,v)${\ plane. It
corresponds to the case where }$u\rightarrow \infty ${. The linearisation
about this critical point gives two eigenvalues: }$\lambda _1=-1${\ with the
eigenvector }$\chi _1=(0,1)${\ and }$\lambda _2=0 ${\ with the eigenvector }$%
\chi _2=(1,1)${\ . Since we have a zero eigenvalue, the stability is
determined by the non-linear behaviour and is one of Lyapunov's 'critical'
cases \cite{sonoda},\cite{bendixon} \cite{bautin}. We apply a linear
transformation to split the system into critical and non-critical variables;
where the critical variables are those eigenvectors with zero eigenvalue and
the non-critical variable are the others. Then we apply a non-linear
transformation which will eliminate the influence of the critical variables
upon the non-critical ones at the leading order.}{\ This gives} $%
(w,v)\rightarrow (W,W+V)${, and the system (\ref{0667non}) becomes: \ }

\begin{eqnarray}
W^{\prime } &=&-2W(W+V)  \label{WV} \\
V^{\prime } &=&-V+2W(W+V)  \nonumber
\end{eqnarray}
{with the critical point at }$(W,V)=(0,0).${The Lyapunov procedure for the
system (\ref{WV}) is to set the linearly stable variable, }$V,${\ equal to
zero in the }$W^{\prime }${\ equation so,}

\[
W^{\prime }=-2W^{2}
\]
{and so the second-order analysis shows that critical point }$(W,V)=(0,0)${\
is unstable. In fact this critical point is a saddle point as can be seen
from the numerical-phase-plane plot figure \ \ref{n23}. The unstable part
correspond to a non-physical range of the cosmological variables, since it
gives $w<0$. The stable part corresponds to the range of values where $w\geq
0$. With respect to our (approximate) exact solution, the unphysical case
will correspond to the range of $x_{0}<0$, since they lead to $\alpha <0$.
Hence the asymptotic solution (\ref{asym}) is the stable late-time behaviour
of the dust universes with small }$\zeta .$ The phase plane structure and
the evolution of $\psi $ vs. $\ln t$ is shown in Figure 5.

\section{Some general asymptotic features}

\subsection{Models with asymptotically constant $\protect\psi $ and $\protect%
\alpha $}

We have seen that $\psi $, and hence the fine structure constant, $\alpha $,
tends to a constant at late times in accelerating universes with power-law
and exponential increase of the scale factor. We can establish a useful
general criterion for this asymptotic behaviour to occur for general $a(t)$.
Suppose that as $t\rightarrow \infty $ both sides of equation (\ref{psi})
tend to a constant (which may be equal to zero). Thus

\begin{eqnarray*}
(\dot{\psi}a^3)^{\cdot } &=&A \\
\psi &=&A\int \frac{tdt}{a^3}+B\int \frac{dt}{a^3}+C\rightarrow C
\end{eqnarray*}
as $t\rightarrow \infty $ if

\begin{eqnarray}
A\int \frac{tdt}{a^3} &\rightarrow &0  \label{lim1} \\
B\int \frac{dt}{a^3} &\rightarrow &0  \label{lim2}
\end{eqnarray}
Then for consistency we also require, as $t\rightarrow \infty $, that

\[
N\exp [-2C]\rightarrow A 
\]
so $A$ cannot be chosen to be zero. Thus in all cosmological models for
which (\ref{lim1}) and (\ref{lim2}) hold there will be a self-consistent
asymptotic solution as $t\rightarrow \infty $ of the form

\[
\psi =N\exp [-2C]\int \frac{tdt}{a^{3}}+B\int \frac{dt}{a^{3}}+C 
\]
where $B,C$ are positive constants. We see that the $n>2/3$ and de Sitter
cases satisfy the conditions (\ref{lim1}) and (\ref{lim2}) hence $\psi $ is
asymptotically constant. The dust ($n=2/3$) case, for which $\psi
\rightarrow \infty $ as $t\rightarrow \infty $ satisfies (\ref{lim2}) but
violates (\ref{lim1}).

\section{Conclusions}

Using a phase plane analysis we have studied the cosmological evolution of a
time-varying fine-structure 'constant' $\alpha (t)=\exp [2\psi ]$, in the
BSBM theory. We have considered the cases created by power-law evolution of
the expansion scale factor of the universe. We have shown that in general $%
\alpha $ increases with time or asymptotes to a constant value at late
times. We have found a new exact solution for the case of a universe
dominated by a stiff fluid or massless scalar field. We have also found
general asymptotic solutions for all the different and possible behaviours
via the analysis of the critical points of the system that determines the
evolution of $\alpha $. In particular, we have found asymptotic solutions
for the dust, radiation and curvature-dominated FRW universe which also
generalise the asymptotes found in \cite{bsm1}.These solutions correspond to
late-time attractors that describes the $\psi $ and $\alpha $ evolution in
time and will enable a more detailed analysis to be made of the fit between
theoretical expectations of varying $\alpha $ theories and observations of
relativistic fine structure in atoms at high redshift.

\vspace{2cm} \noindent \textbf{Acknowledgements }We would like to thank%
\textbf{\ }H\aa vard Sandvik and Jo\~{a}o Magueijo for discussions. DFM is
supported by Funda\c{c}\~ao para a Ci\^encia e a Tecnologia, Portugal,
through the research grant BD/15981/98.

\vspace{2cm} \noindent \appendix\textbf{\appendixname :} \vspace{.5cm}
\noindent \textbf{\ General analysis of the phase plane bifurcations}

In previous sections we have analysed the $\psi $ evolution equation (\ref
{psi}) for a range of variables which are physically realistic and
correspond to expanding universes. We will now analyse the whole range for
variables of the system (\ref{llog}). As before we see there are two
critical points in the $(w,v)$ plane, at $(0,-1+{\frac{3n}{2}})$ and $%
((1-3n)(\frac{3n}{2}-1),0)$. Linearising (\ref{llog}) about $%
(w_{c_{1}},v_{c_{1}})=(0,-1+\frac{3n}{2})$ and $%
(w_{c_{2}},v_{c_{2}})=((1-3n)(\frac{3n}{2}-1),0)$ we obtain the following
characteristics matrices:

$M_1=%
\bordermatrix{ & &\cr
		 & 2-3n  & 0   \cr
                 & 1 & 1-3n \cr} $ \qquad $M_2=%
\bordermatrix{& & \cr
           & 0 & (3n-1)(2-3n)    \cr
           & 1 & 1-3n    \cr} $

The characteristic matrices are non singular except when $n=\frac{1}{3}$ or $%
n=\frac{2}{3}$. In the non-singular cases the critical points will be simple
and the system defined by these differential equations is structurally
stable \cite{andronov}, and there will be no 'strange' chaotic behaviour
outside the neighbourhood of the critical points. Hence, the linearised
system will have the same phase portrait as non-linearised one in the
neighbourhood of the critical points.

The evolution, with respect to changing $n$, of the signs of the determinant
and the trace of these two matrices is given in the table. This show us that
there are always two critical points in our system, an unstable saddle and
an attractor (which changes from a spiral to a node).

\vspace{.5cm}

\begin{tabular}[hb]{|c|c|c|}
\hline
\textbf{$n$} & \multicolumn{2}{|c|}{\textbf{Critical Points $(w_{c},v_{c})$}}
\\ \cline{2-3}
& $\left( \left( 3n-1\right) \left( 1-\frac{3n}{2}\right) ,0\right) $ & $%
\left( 0,1-\frac{3n}{2}\right) $ \\ \hline
$(-\infty ;\frac{1}{3})$ & Saddle Point (non-physical) & Unstable Node \\ 
& det $M_{1}<0$, Tr $M_{1}>0$ & det $M_{2}>0$, Tr $M_{2}>0$ \\ \hline\hline
$\frac{1}{3}$ & Origin & Axis \\ 
& det $M_{1}=0$, Tr $M_{1}=0$ & det $M_{2}=0$, Tr $M_{2}>0$ \\ \hline\hline
$(\frac{1}{3};\frac{1}{2})$ & Stable Spiral & Saddle Point \\ 
& det $M_{1}>0$, Tr $M_{1}<0$ & det $M_{2}<0$, Tr $M_{2}>0$ \\ \hline\hline
$\frac{1}{2}$ & Stable Spiral & Saddle Point \\ 
& det $M_{1}>0$, Tr $M_{1}<0$ & det $M_{2}<0$, Tr $M_{2}=0$ \\ \hline\hline
$(\frac{1}{2};\frac{3}{5})$ & Stable Spiral & Saddle Point \\ 
& det $M_{1}>0$, Tr $M_{1}<0$ & det $M_{2}<0$, Tr $M_{2}<0$ \\ \hline\hline
$\frac{3}{5}$ & Stable Spiral (node) & Saddle Point \\ 
& det $M_{1}>0$, Tr $M_{1}<0$ & det $M_{2}<0$, Tr $M_{2}<0$ \\ \hline\hline
$(\frac{3}{5};\frac{2}{3})$ & Stable Node & Saddle Point \\ 
& det $M_{1}>0$, Tr $M_{1}<0$ & det $M_{2}<0$, Tr $M_{2}<0$ \\ \hline\hline
$\frac{2}{3}$ & Stable Axis & Stable Axis \\ 
& det $M_{1}=0$, Tr $M_{1}<0$ & det $M_{2}=0$, Tr $M_{2}<0$ \\ \hline\hline
$(\frac{2}{3};\infty )$ & Saddle Point (non-physical) & Stable Node \\ 
& det $M_{1}<0$, Tr $M_{1}<0$ & det $M_{2}>0$, Tr $M_{2}<0$ \\ \hline
\end{tabular}

\vspace{.5cm}

\begin{figure}[hb]
\centering
\epsfig{file=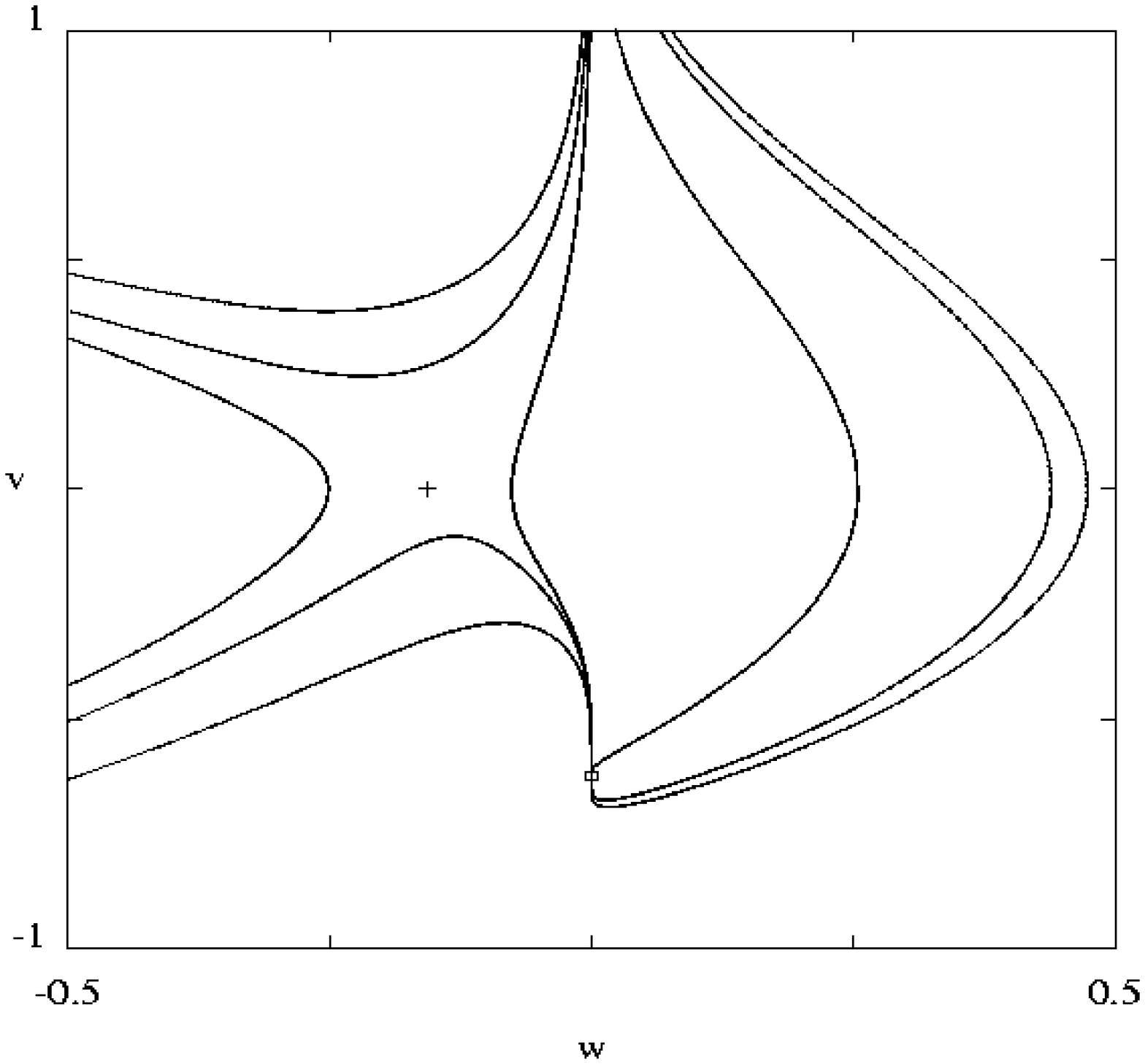,height=6cm} %
\epsfig{file=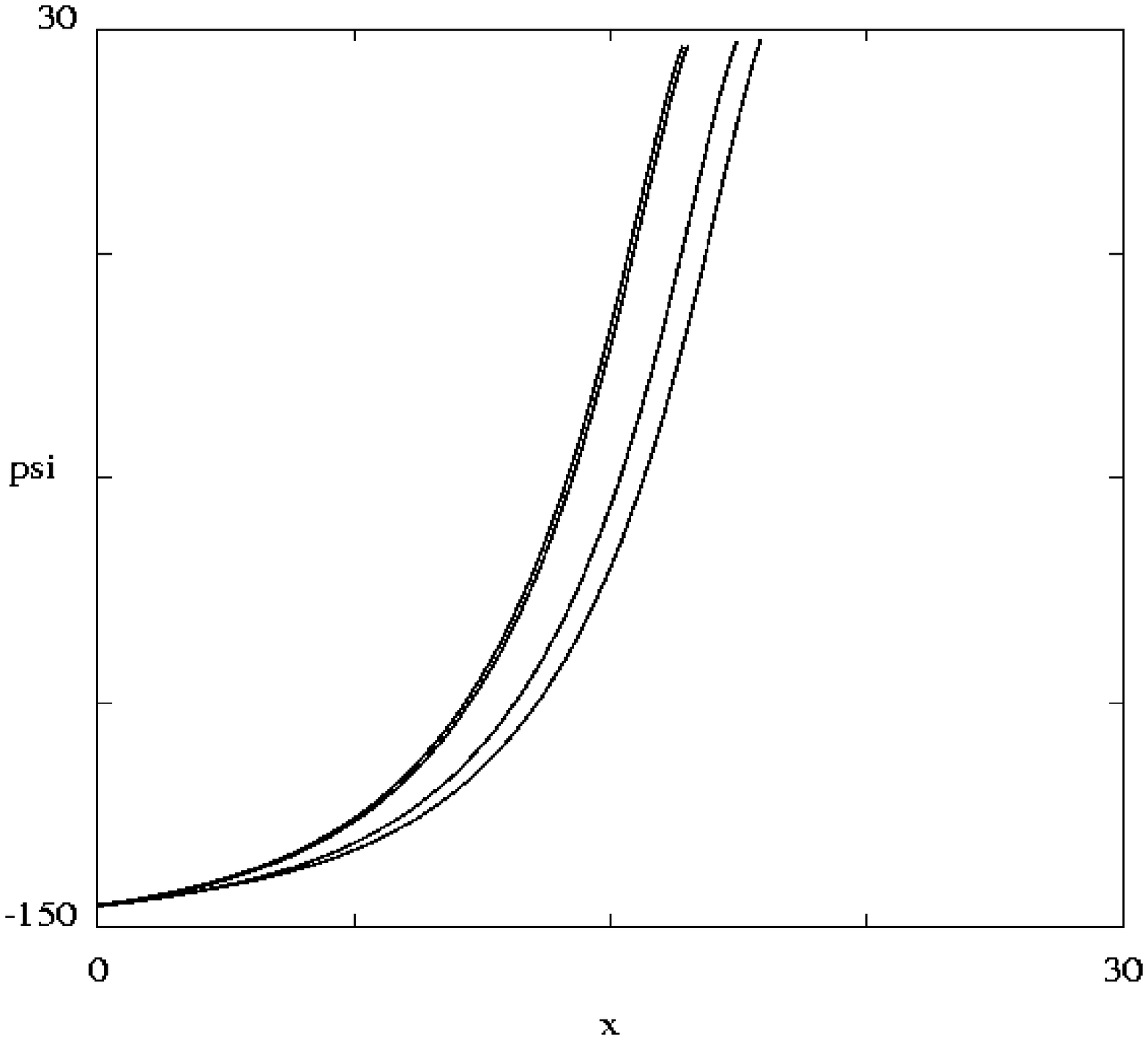,height=6cm}
\caption{{\protect\small \textit{Numerical plots of the phase space in the $%
(w,v)$ coordinates, and the $\protect\psi$ evolution with $x=log(t)$
for $n=\frac{1}{4}$. The '$+$' sign is a saddle point and the square
is an unstable source. }}} 
\label{n025}
\end{figure}

The cases $n=\frac{1}{3}$ or $n=\frac{2}{3}$, where the determinant of the
characteristic matrixes vanishes, lead to a bifurcation of codimension $1$,
in particular, of Saddle-Node type \cite{wiggins}, since they correspond to
points where the determinants of the characteristic matrices change sign, det%
$M_{1}$ $=$ det$M_{2}=0$. At these values of $n$ the nature of the system
will change. Cosmologically, these points represent a change in the
behaviour of the time evolution of the fine structure 'constant' as can be
seen from the figures: \ref{n025}, \ref{n13}, \ref{n12}, \ref{n35}, \ref{n23}%
, \ref{n1}, which display the time evolution of $\psi$. When $n$\ starts to
grows from $-\infty $\ to $\frac{1}{3}$\ the two critical points slowly\emph{%
\ }converge at $n=\frac{1}{3}$\emph{. }For example, in the $n=0$ case where
we may without loss of generality set $a=1$, equation (\ref{psi}) has the
exact solution

\[
\alpha =\exp [2\psi ]=A^{-2}\cosh ^{2}[AN^{1/2}(t+t_{0})]
\]
where $A,t_{0}$ are constants. This is an unrealistically rapid growth
asymptotically , $\psi \varpropto t,$caused by the absence of the inhibiting
effect of the cosmological expansion. The case of $n=1/4$ is shown in Figure
6, which shows the phase space trajectories and the evolution of $\psi $ vs. 
$\ln t.$ 

At $n=\frac{1}{3}$ we are in the situation where the two critical points
collapse into a unique one at the origin, creating a saddle-node bifurcation
and a concomitant change in the behaviour and evolution of $\psi $. As $n$
keeps growing the single critical point splits into two critical points
again. They move apart until the radiation value is reached, $n=\frac{1}{2}$
(Tr $M_{2}=0$). In this case, $\psi $ is a asymptotically monotonic growing
function of time, with some small oscillations near the Planck epoch.
However, note that in our universe the asymptote giving an increase of $\psi 
$ behaviour with time is never reached before the dust-dominated evolution
takes over \cite{bsbm}, \cite{bsm1}, \cite{bsm2}.

As the universe evolves to the dust-dominated epoch, and $n$ approaches the
intermediate behaviour $n=\frac{3}{5}$, the two critical points start to
coalesce again into a single point. When $n=\frac{3}{5}$ is reached, $\psi $
becomes a strictly monotonically growing function of time. When $n$ reaches
the value corresponding to a dust-dominated universe, $n=\frac{2}{3}$,
another saddle-node bifurcation occurs. The two critical points collapse
into a single one. Again there will be change in the behaviour of $\psi $
for larger values of $n$. Accordingly, when $n>2/3,$ the two critical points
reappear once again and $\psi $ becomes asymptotically constant in value.

Notice, that although a bifurcation is something that 'spoils' the smooth
behaviour of a system, in our case, that won't happen, due to the physical
constraints of our variables. In reality due to those constraints, the
physical system will never 'feel' the abrupt change at $n=\frac{1}{3}$ and $%
n=\frac{2}{3}$. This is also due to the fact that the attracting critical
point always lies in the physical range of the variables, while the unstable
one disappears form the physical system when the bifurcations occur, as can
be seen from the phase plane plots.

\end{document}